\documentclass[
  aps,                  
  prd,                  
  reprint,              
  nofootinbib,          
  superscriptaddress,   
  showpacs,             
  showkeys,             
  preprintnumbers,      
  longbibliography      
]{revtex4-1}

\usepackage[utf8]{inputenc}
\usepackage{amsfonts,amsmath,amssymb}
\usepackage{bm,dsfont} 
\usepackage{url}
\usepackage{graphicx}
\usepackage{hyperref}
\hypersetup{linktocpage,colorlinks=true,urlcolor=purple,linkcolor=purple,citecolor=purple}
\usepackage{multirow}
\usepackage{xcolor}

\newcommand{\udec}{Departamento de Fí­sica, Universidad de Concepción, Casilla 160-C, Concepción, Chile}
\newcommand{\pucv}{Instituto de Física, Pontificia Universidad Católica de Valparaíso, Avenida Brasil 2950, Valparaíso, Chile}
\newcommand{\unap}{Instituto de Ciencias Exactas y Naturales (ICEN), Facultad de Ciencias, Universidad Arturo Prat, Iquique, Chile}

\begin{document}

\title{The Spin Tensor of Dark Matter and the Hubble Parameter Tension}

\author{Fernando Izaurieta}
\email{fizaurie@udec.cl}
\affiliation{\udec}

\author{Samuel Lepe}
\email{samuel.lepe@pucv.cl}
\affiliation{\pucv}

\author{Omar Valdivia}
\email{ovaldivi@unap.cl}
\affiliation{\unap}

\date{\today}

\begin{abstract}
  Allowing for a nonvanishing spin tensor for cold dark matter ($\omega_{\mathrm{DM}}=0$) has the consequence of giving rise to an effective $\mathrm{FLRW}$ dynamics with a small negative barotropic constant for an effective dark matter density ($-1/3<\omega_{\mathrm{eff}}\leq0$). This turns out to solve the Hubble parameter tension in a straightforward way.
\end{abstract}



\maketitle

%
%

\section{Introduction}

The Hubble parameter tension has given rise to a considerable number of hypotheses to explain it (see Refs.~\cite{Lucca:2020zjb,Alestas:2020mvb,Vattis:2019efj,Guo:2018ans,DiValentino:2017iww,Rossi:2019lgt,Braglia:2020iik,Alcaniz:2019kah,Jedamzik:2020krr}). They range from possible systematic measurement errors (See Ref.~\cite{Freedman:2020dne}), to modified dark energy~\cite{Alestas:2020mvb,Zumalacarregui:2020cjh,Yang:2018euj,Vagnozzi:2019ezj,DiValentino:2019jae}, to more exotic theories as nonminimal couplings, torsional topological invariants, or quadratic Poincar\'{e} Gauge Theory\footnote{For a complete description of PGT's, see~\cite{Hehl1980,Blagojevic:2013xpa}. Other recent developments related to torsion can be found in \cite{Chakrabarty:2018ybk}} (See Refs.~\cite{Alexander:2019wne,Magueijo:2019vmk,Barker:2020gcp}), among many others. An independent improvement in the measurement of $H_{0}$ can be expected in the future using black hole mergers as dark standard sirens, see Ref.~\cite{Soares-Santos:2019irc}.

Among all this buzz of activity, Ref.~\cite{Poplawski:2019tub} offered a particularly simple solution: dropping the coldness hypothesis in dark matter and allowing for a small negative value for the barotropic constant ($\omega_{\mathrm{DM}}=-0.0108$) fixes the Hubble parameter tension. Some arguments may favor such a non-particle dark matter scheme (see Ref.~\cite{Poplawski:2013mra}), but it would seem like an exotic possibility for many.

This article offers an alternative: a nonvanishing spin tensor for cold dark matter ($\omega_{\mathrm{DM}}=0$) may solve the problem along the same lines presented in Ref.~\cite{Poplawski:2019tub}. The spin tensor of cold dark matter gives rise to an effective dynamic in the FLRW equations corresponding to a small and negative effective barotropic constant $-1/3<\omega_{\mathrm{eff}}\leq0$, precisely as required in Ref.~\cite{Poplawski:2019tub} to fix the Hubble tension.

Studying a nonvanishing spin tensor also opens the possibility of a nonvanishing torsion. Many well-motivated Lagrangian choices deal with torsion and lead to different generalized gravity theories. However, in this work, we choose a theory as close to General Relativity (GR) as possible. It is just the standard Einstein-Hilbert term, the cosmological constant, and minimally coupled matter in an \textquotedblleft a la Palatini\textquotedblright\ approach, i.e., Einstein-Cartan gravity.

The meagerness of our choice should not be considered demeaning to more general theories. On the contrary, there are many good reasons to look far beyond the Einstein-Hilbert term and standard minimal couplings, as gauge invariance~\cite{Hehl1980,Blagojevic:2013xpa}, propagating torsion~\cite{Carroll:1994dq}, or modeling neutrino oscillations~\cite{Chakrabarty:2019cau}. Despite this, we choose to work with the Einstein-Cartan theory on purpose. The reason for it is its simplicity. We can find many theories of modified torsional gravity and new particles that may explain the current issues in cosmology in the literature. However, they substantially depart from the General Relativity and the Standard Model Lagrangians. Considering the enormous experimental success of both theories, we think it is worth exploring the opposite approach. It corresponds to explore whether a small modification of General Relativity may suffice to explain these issues, without requiring any extra couplings with the matter Lagrangian, or additional torsional terms.

For the particular case of the Hubble parameter tension, the answer to this question is yes. We found that to lift a seemingly unimportant hypothesis (vanishing spin tensor for cold dark matter) explains the Hubble parameter tension without requiring terms beyond Einstein-Hilbert, and leaving the $\Lambda\mathrm{CDM}$ dynamics almost unchanged. What makes the result remarkable is precisely the modesty of the change.

\section{The spin tensor of cold, non-interacting dark matter}

Our knowledge of dark matter nature is scarce, and it has given rise to a vast number of models about its composition. This article does not add a new model to this enormous collection. The goal of this section is to deduce an Ansatz for the effective spin tensor of dark matter at a cosmological scale. To do this, we will only consider cold, non-interacting dark matter, the Copernican cosmological symmetries, and dimensional consistency arguments. We advance no further hypotheses on the nature of the particles that may constitute dark matter itself.




Let us briefly revisit the spin tensor\footnote{The spin tensor corresponds to the variation of the matter Lagrangian with respect to the affine structure. It depends on the intrinsic quantum mechanical spin of the particles that are the source of the gravitational field, and should not be confused with the angular momentum density of a rotating body.} role as a source of gravity in the simplest possible scenario. The spin tensor's contribution appears when considering an \textquotedblleft a la Palatini\textquotedblright\ approach, i.e., with the connection and the metric as independent degrees of freedom, and without imposing the torsionless condition (Riemann-Cartan geometry)~\cite{Kib61,Sciama:1964wt,Hehl:1971qi,doi:10.1142/6742,doi:10.1142/0356,Hehl76,Shapiro:2001rz,Hammond:2002rm,Poplawski:2009fb}.

On this geometry, let us consider the usual Einstein-Hilbert and cosmological constant Lagrangian with minimally-coupled matter $\mathcal{L}=\frac{1}{2\kappa_{\mathrm{4}}}\left[  R\left(  \Gamma,g\right)  -2\Lambda \right]+\mathcal{L}_{\mathrm{M}}$  (Einstein-Cartan gravity). The corresponding field equations read
\begin{align}
&R_{\mu \nu}-\frac{1}{2}g_{\mu \nu}R+\Lambda g_{\mu \nu}   =\kappa_{\mathrm{4}}\tau_{\mu \nu}\,,\label{Eq_metric}\\
 &T^{\lambda}{}_{\mu \nu}-\delta_{\mu}^{\lambda}T^{\gamma}{}_{\gamma \nu}+\delta_{\nu}^{\lambda}T^{\gamma}{}_{\gamma \mu}  =\kappa_{\mathrm{4}}\sigma^{\lambda}{}_{\mu \nu}\,.\label{Eq_affine}
\end{align}

Here $\kappa_{\mathrm{4}}=\frac{8\pi G}{c^{4}}$, and $R_{\mu \nu}$ and $R$ correspond to the Ricci tensor and Ricci scalar constructed from the general connection $\Gamma_{\mu \nu}^{\lambda}$ (and not the Christoffel). The $\sigma^{\lambda}{}_{\mu \nu}$ stands for the spin tensor of $\mathcal{L}_{\mathrm{M}}$, $\tau_{\mu \nu}$ for its stress-energy tensor, and $T^{\lambda}{}_{\mu \nu}=\Gamma_{\mu \nu}^{\lambda}-\Gamma_{\nu \mu}^{\lambda}$ for torsion. 

Tracing Eq.~(\ref{Eq_affine}), the torsion reads
\begin{equation}
T^{\lambda}{}_{\mu \nu}=\kappa_{\mathrm{4}}\left[  \sigma^{\lambda}{}_{\mu \nu}+\frac{1}{2}\left(  \sigma^{\gamma}{}_{\gamma \mu}\delta_{\nu}^{\lambda}-\sigma^{\gamma}{}_{\gamma \nu}\delta_{\mu}^{\lambda}\right)  \right]\,.\label{Eq_Torsion_desp}
\end{equation}
The Eq.~(\ref{Eq_Torsion_desp}) implies that in Einstein-Cartan theory, torsion does not propagate in a vacuum\footnote{Of course, this is only in the context of the closest alternative to GR, the Einstein-Cartan theory. When considering more general Lagrangians, with nonminimal couplings or quadratic terms in curvature and torsion (See Refs.~\cite{Carroll:1994dq,Alexander:2019wne,Magueijo:2019vmk,Barker:2020gcp,Valdivia:2017sat,Cid:2017wtf,Barrientos:2019awg,Alexander:2020umk}), torsion propagates. For more information on the wave operator on spaces with Riemann-Cartan geometry and propagation of perturbations, see
Refs.\cite{Barrientos:2019msu,Barrientos:2019awg}.}, in sharp contrast with curvature. Even further, in the case of standard baryonic matter, the spin tensor arises as a small quantum effect. Both things together imply that in almost any normal astrophysical situation, the torsion created trough this mechanism is so negligible\footnote{Again, it is a weak four-fermion interaction effect only in the case of Einstein-Cartan gravity. More complex theories can produce more significant torsional effects that propagate in a vacuum, see Ref.~\cite{Carroll:1994dq,Puetzfeld:2014sja}. Also, torsion may have some effect on neutrino oscillations, see Ref.~\cite{Chakrabarty:2019cau}} that it is possible to make light-hearted jokes about it (See the end of Chap.~8.4 of Ref.~\cite{SupergravityVanProeyen}).

Therefore, in the context of highly-localized interacting baryonic matter and the Einstein-Hilbert Lagrangian, torsion seems negligible, and the Einstein-Cartan theory reduces to standard General Relativity. When considering a cosmological scale, torsion seems even more negligible: since it does not propagate in a vacuum, the effective spin tensor of a cosmological \textquotedblleft gas of galaxies\textquotedblright \ clearly vanishes.
That is why, in order to have torsion propagation, in the literature there are more complex Lagrangians and matter couplings~\cite{Carroll:1994dq,Hehl1980,Blagojevic:2013xpa}. However, in this article, we will stick to the most straightforward, non-propagating Einstein-Cartan phenomenology of Eqs.~(\ref{Eq_metric}-\ref{Eq_Torsion_desp}) from now on.

Let us observe that the full description we have made depends on a simple fact: baryons interact. Their interactions give rise to decoherence and localized structure, from atoms to stars, and huge volumes devoided of them. That is why non-propagating features of baryons (as their spin tensor and torsion) are irrelevant in cosmological scales in the late evolution.

The case of non-interacting cold dark matter is precisely the opposite. Whatever it is, cold dark matter characterizes for the lack of interaction with Standard Model particles and itself. Therefore, independently of its composition, it seems reasonable to expect extremely weak or absent decoherence effects for its particles. In this case, we would have extremely delocalized dark matter particles, with giant wave functions extending over vast distances in the Universe. It is in sharp contrast with the highly-localized baryons.

With no decoherence effects and intergalactic-wide dark matter wave functions, quantum effects as the spin tensor may be much more relevant. According to Eq.~(\ref{Eq_Torsion_desp}), the spin tensor can create a nonvanishing torsion at cosmic scales in this case. This torsion would contribute to cosmological evolution, even if it cannot propagate in a vacuum. A thin but nonvanishing dark matter wave amplitude on cosmological scales can support an equally widely distributed spin tensor and torsion.

The problem is how to model the spin tensor of dark matter.  Again, some reasonable models for baryonic matter do not seem to apply to dark matter. For instance, the usual way to model the spin tensor of highly-interacting baryonic matter is as a Weyssenhof fluid~\cite{Weyssenhoff:1947iua,Obukhov:1987yu}.


The difficulties with Weyssenhof fluids start by the fact that their spin tensors do not obey isotropy~\cite{Boehmer:2006gd} in the short micro-scales of a highly interacting fluid. However, on a macro-scale, all the local anisotropies can be averaged, recovering the Copernican principle. The result is a theory like the ones of Refs.~\cite{Brechet:2008zz,Poplawski:2011jz,Poplawski:2010kb,Poplawski:2014dea,Poplawski:2018ypb}, where the spin tensor of standard model fermions gives rise to a Big-Bounce model under the extreme high-densities of very early times.

The same reasoning does not seem to apply to non-interacting cold dark matter on broad cosmic scales, and a spin tensor obeying the Copernican principle seems more appropriate. Imposing the Copernican principle on the spin tensor $\pounds _{\zeta}\sigma_{\lambda \mu \nu}=0$, we get something of the form,
\begin{align}
\sigma_{\lambda \mu \nu}  & =\frac{2}{c^{2}\kappa_{\mathrm{4}}}X\left( t\right)  \left(  g_{\lambda \mu}U_{\nu}-g_{\lambda \nu}U_{\mu}\right)
+\nonumber \\
& -\frac{2}{c^{2}\kappa_{\mathrm{4}}}Y\left(  t\right)  \sqrt{\left \vert g\right \vert }\epsilon_{\lambda \mu \nu \rho}U^{\rho}\,  , \label{Eq_Copernican_Spin}%
\end{align}
where $U^{\rho}$ is the co-moving 4-velocity, and $X\left(  t\right)  $ and $Y\left(  t\right)  $ are arbitrary functions of time. Eq.~(\ref{Eq_Copernican_Spin}) is a well-known Ansatz; the second term sometimes receives the nickname of Cartan's staircase, see Ref.~\cite{Alexander:2020umk}.

Since torsion depends algebraically on the spin tensor through Eq.~(\ref{Eq_Torsion_desp}), it corresponds to
\begin{equation}
T_{\lambda \mu \nu}=\frac{1}{c^{2}}\left[  X\left(  g_{\lambda \nu}g_{\mu \rho }-g_{\lambda \mu}g_{\nu \rho}\right)  -2\sqrt{\left \vert g\right \vert }Y\epsilon_{\lambda \mu \nu \rho}\right] U^{\rho}\,.\label{Eq_Torsion_XY}
\end{equation}

Without a dark matter theory, we have no a priori information on $X\left( t\right)  $ or $Y\left(  t\right)  $. To deal with this, we have two alternatives. The first alternative would be to look for a fundamental theory of dark matter particles that can give rise to a spin tensor with the form Eq.~(\ref{Eq_Copernican_Spin}). One example of such fundamental hypothetical particles would be dark spinors, Ref.~\cite{Boehmer:2008ah}. They are dark matter candidates giving rise to a spin tensor compatible with homogeneity and isotropy. However,  even if dark spinors provide an excellent solution, it is not clear that it is unique. There could be other (maybe unknown) dark matter candidates that also provide such spin tensor. That is why we decided to use a more general approach and to leave that door open.
This second approach is to deduce a torsional equation of state without resorting to a particular dark matter Lagrangian. To do so, let us observe that putting all the torsional terms of $R_{\mu \nu}$ and $R$ at the right-hand side of Eq.~(\ref{Eq_metric}), we get
\begin{equation}
\mathring{R}_{\mu \nu}-\frac{1}{2}g_{\mu \nu}\mathring{R}+\Lambda g_{\mu \nu}=\tau_{\mu \nu}+\tau_{\mu \nu}^{\left(  \mathrm{T}\right)  } ,\label{Eq_TorsionRightHandSide}
\end{equation}
where $\mathring{R}_{\mu \nu}$ and $\mathring{R}$ are the standard torsionless Ricci tensor and scalar, and $\tau_{\mu \nu}^{\left(  \mathrm{T}\right)  }$ is the effective stress-energy tensor for torsion. It corresponds to
\begin{align}
\tau_{\mu \nu}^{\left(  \mathrm{T}\right)  }  & =g_{\mu \nu}\left( \mathring{\nabla}_{\alpha}K^{\alpha \rho}{}_{\rho}+\frac{1}{2}\left[ K^{\alpha}{}_{\lambda \alpha}K^{\lambda \rho}{}_{\rho}-K^{\alpha}{}_{\lambda \rho}K^{\lambda \rho}{}_{\alpha}\right]  \right)  +\nonumber \\
& +\frac{1}{2}\left(  \mathring{\nabla}_{\nu}K^{\alpha}{}_{\mu \alpha }+\mathring{\nabla}_{\mu}K^{\alpha}{}_{\nu \alpha}+K^{\alpha}{}_{\lambda \mu }K^{\lambda}{}_{\nu \alpha}+\right.  \nonumber \\
& +\left.  K^{\alpha}{}_{\lambda \nu}K^{\lambda}{}_{\mu \alpha}-\left[ \mathring{\nabla}_{\lambda}+K^{\alpha}{}_{\lambda \alpha}\right]  \left[
K^{\lambda}{}_{\mu \nu}+K^{\lambda}{}_{\nu \mu}\right]  \right)\label{Eq_T_eff_torsion}
\end{align}
where
\begin{equation}
K_{\mu \nu \lambda}=\frac{1}{2}\left(  T_{\nu \mu \lambda}-T_{\mu \nu \lambda}+T_{\lambda \mu \nu}\right)
\end{equation}
corresponds to what is called the contorsion (or contortion) tensor.

Since this effective stress-energy tensor Eq.~(\ref{Eq_T_eff_torsion}) is quadratic in torsion, it indicates that torsion scales as
\begin{equation}
T_{\lambda \mu \nu}\sim \sqrt{\mathrm{energy~density}} .
\end{equation}
Therefore, it is reasonable to expect $X\left(  t\right)  $ and $Y\left( t\right)  $ to scale in the same way. Even further, torsion has to vanish in a vacuum, and it seems natural to expect it to grow for higher dark matter densities. Since in the Lagrangian, the only relevant energy densities are the cold dark matter density $\rho_{\mathrm{DM}}$ and $\frac{\Lambda}{\kappa_{\mathrm{4}}}$, it is natural to consider an Ansatz for the torsional equation of state as
\begin{equation}
X,Y\propto \left(  \frac{\kappa_{\mathrm{4}}\rho_{\mathrm{DM}}}{\Lambda}\right)^{N}\sqrt{\rho_{\mathrm{DM}}}\,,\label{Eq_ansatz_XY}
\end{equation}
where $N>-1/2$ is some number (in general, a different one for $X$ and $Y$). To assume there is no relation between torsion and $\Lambda$ amounts to the choice $N=0$.

Let us observe what is happening. Copernican symmetries constraint the spin tensor so strongly in the cosmological scale, that it is possible to use dimensional consistency arguments to fix the dependency of $X$ and $Y$ on the dark matter density $\rho_{\mathrm{DM}}$. Finally, it leads us to a torsional equation of state in terms of an adimensional \textquotedblleft torsiotropic\textquotedblright\ constant $\alpha_{\mathrm{Y}}$, see Eq.~(\ref{Eq_Torsiotropic}). This way, we have done for the spin tensor the same thing we usually do for the stress-energy tensor on an FLRW geometry. Whatever the cosmological fluid is, we describe it through $p=\omega\rho$ and choose different barotropic constants $\omega$ depending on the fluid. The spin tensor components may not scale linearly with the dark matter density, but the concept is the same.


\section{The FLRW equations}

Using the FLRW metric with flat spatial section and the Eq.~(\ref{Eq_Torsion_XY}), the field equation~(\ref{Eq_metric}) takes the form
\begin{equation}
\frac{3}{c^{2}}\left[  \left(  H+X\right)  ^{2}-Y^{2}\right]  -\Lambda=\kappa_{\mathrm{4}}\left(  \rho_{\mathrm{b}}+\rho_{\mathrm{DM}}\right)\,,
\end{equation}
and
\begin{align}
\left(  \frac{1}{c^{2}}\left(  \dot{H}+\dot{X}\right)  +H\left(  H+X\right)+\right. &  \nonumber \\
\left.  +\frac{1}{2}\left[  \left(  H+X\right)  ^{2}-Y^{2}\right]  \right)-\frac{1}{2}\Lambda & =-\kappa_{\mathrm{4}}\frac{1}{2}\left(  p_{\mathrm{b}}+p_{\mathrm{DM}}\right) ,
\end{align}
where $\rho_{\mathrm{b}}$ and $p_{\mathrm{b}}$ are the density and pressure of baryons. Afterward, we impose the cold dark matter pressureless condition $p_{\mathrm{DM}}=0$, but for now, let us keep the term so we can follow its role in the equations.

Let us impose the condition that baryonic matter does not directly interchange energy with dark matter and neither with torsion. The former is an empirical fact, and the second is, to the best of our knowledge, an excellent approximation (except for very early times). This lack of interaction with baryonic matter implies the usual conservation law for it
\begin{equation}
\frac{\mathrm{d}}{\mathrm{d}t}\rho_{\mathrm{b}}+3H\left(  \rho_{\mathrm{b}}+p_{\mathrm{b}}\right)  =0 \, .
\end{equation}

From this (and some straightforward manipulation), it is possible to write the field equations as
\begin{align}
&\frac{3}{c^{2}}\left(  H+X\right)  ^{2}-\Lambda -\kappa_{\mathrm{4}}\left( \rho_{\mathrm{b}}+\rho_{\mathrm{DM}}+\frac{3Y^{2}}{\kappa_{\mathrm{4}}c^{2}}\right)=0\,,\\
&\frac{\mathrm{d}}{\mathrm{d}t}\rho_{\mathrm{b}}+3H\left(  \rho_{\mathrm{b}}+p_{\mathrm{b}}\right)     =0\,,\\\nonumber
&\frac{\mathrm{d}}{\mathrm{d}t}\left(  \rho_{\mathrm{DM}}+\frac{3Y^{2}}{\kappa_{\mathrm{4}}c^{2}}\right)  +3H\left(  \rho_{\mathrm{DM}}+\frac
{2Y^{2}}{\kappa_{\mathrm{4}}c^{2}}+p_{\mathrm{DM}}\right)  + \\
&+X\left(  \rho_{\mathrm{b}}+\rho_{\mathrm{DM}}+3\left[  p_{\mathrm{b}}+p_{\mathrm{DM}}\right]  -2\frac{\Lambda}{\kappa_{\mathrm{4}}}\right) 
=0\,.
\end{align}

The roles of the two spin tensor modes are extremely different. Let us sacrifice generality for simplicity, and let us impose $X=0$ to focus our
attention only on the behavior of the $Y$ component. Defining an effective dark matter density and pressure given by
\begin{align}
\rho_{\mathrm{eff}} &  =\rho_{\mathrm{DM}}+\frac{3}{c^{2}\kappa_{\mathrm{4}}}Y^{2}\,,\label{Eq_dens_eff}\\
p_{\mathrm{eff}} &  =p_{\mathrm{DM}}-\frac{1}{c^{2}\kappa_{\mathrm{4}}}Y^{2}\,,\label{Eq_press_eff}
\end{align}
in the $X=0$ case, the field equations adopt the canonical form
\begin{align}
&\frac{3}{c^{2}}H^{2}   =\Lambda+\kappa_{\mathrm{4}}\left(  \rho_{\mathrm{b}}+\rho_{\mathrm{eff}}\right)\,,\\
&\frac{\mathrm{d}}{\mathrm{d}t}\rho_{\mathrm{b}}+3H\left(  \rho_{\mathrm{b}}+p_{\mathrm{b}}\right)     =0\,,\\
&\frac{\mathrm{d}}{\mathrm{d}t}\rho_{\mathrm{eff}}+3H\left(  \rho_{\mathrm{eff}}+p_{\mathrm{eff}}\right)     =0\,.
\end{align}

The only difference with the standard $\mathrm{\Lambda CDM}$ cosmology is that the  effective pressure is still negative $p_{\mathrm{eff}}<0$ for the case of cold dark matter $p_{\mathrm{DM}}=0$. Let us use the Ansatz~(\ref{Eq_ansatz_XY}) with $N=0$ to write down
\begin{equation}
Y=\alpha_{\mathrm{Y}}\, c \,\sqrt{\frac{\kappa_{\mathrm{4}}}{3}\rho_{\mathrm{DM}}}\,,\label{Eq_Torsiotropic}
\end{equation}
where $\alpha_{\mathrm{Y}}$ is a \textquotedblleft torsiotropic\textquotedblright\ constant of proportionality.

In the case of cold dark matter $p_{\mathrm{DM}}=0$, the effective barotropic constant corresponds to
\begin{equation}
\omega_{\mathrm{eff}}=\frac{p_{\mathrm{eff}}}{\rho_{\mathrm{eff}}}=-\frac{1}{3}\frac{1}{1+1/\alpha_{\mathrm{Y}}^{2}}\,,\label{Eq_w_eff}
\end{equation}
and therefore $-1/3<\omega_{\mathrm{eff}}\leq0$. In other words, the spin tensor component has the effect of creating a non-particle FLRW effective behavior for otherwise standard cold dark matter. Let us emphasize that this small negative pressure $p_{\mathrm{eff}}$ is only an effective artifact at the FLRW evolution level, and it does not change the characteristic dark matter speed. Dark matter in our model is still standard $\omega_{\mathrm{DM}}=0$ pressureless cold dark matter. The only difference with the standard case is that we assume one of the spin tensor components does not vanish.

Even further, torsion does not interact with Standard Model bosons, and the interaction with fermions is too weak to be detected in a particle physics experiment\footnote{Again, assuming a minimal coupling in Einstein-Cartan theory. There is no experimental reason to assume otherwise, but if there were nonminimal couplings, torsion would be easier to detect, see Ref.~\cite{Puetzfeld:2014sja}.}. Besides, when passing all the torsional terms at the right-hand side of the field equations as in Eq.~(\ref{Eq_TorsionRightHandSide}), it behaves as an additional source of standard torsionless Riemannian gravity. For this reason, some authors have suggested that the whole dark matter phenomenology could be due to torsion~\cite{Tilquin:2011bu}. We do not go as far, but the point is that torsion behaves as an extra dark source for the torsionless piece of gravity.


Therefore, in the current context, we have to distinguish between dressed and bare dark matter. When we measure torsionless Riemannian effects and attribute them to dark matter, what we are measuring is the altogether combined effect of \textquotedblleft bare\textquotedblright\ dark matter and torsion, $\rho_{\mathrm{DM}}+\frac{3}{c^{2}\kappa_{\mathrm{4}}}Y^{2}$. In other words, torsion amplifies the effect of dark matter, creating the \textquotedblleft dressed\textquotedblright\ version $\rho_{\mathrm{eff}}$ we observe as a dark source.

The solution to the Hubble parameter tension presented in Ref.~\cite{Poplawski:2019tub}, but rephrased in terms of \textquotedblleft dressed\textquotedblright \ dark matter, would be the following. According to the PLANCK report Ref.~\cite{Aghanim:2018eyx}, they measured in a
model-independent way the combination $\Omega_{x}h^{2}$ for baryons and dark matter (where $h$ is the \textquotedblleft adimensional\textquotedblright\ Hubble parameter given by $H_{0}=100h\, \frac{\mathrm{km}}{\mathrm{s}\cdot \mathrm{Mpc}}$) as
\begin{align}
\Omega_{\mathrm{b}}^{\left(  \mathrm{PLANCK}\right)  }h^{2} &  =0.0224\pm 0.0001\,,\label{Eq_Omega_b_PLANCK}\\
\Omega_{\mathrm{DM}}^{\left(  \mathrm{PLANCK}\right)  }h^{2} &  =0.120\pm 0.001\,.\label{Eq_Omega_DM_PLANCK}
\end{align}

The tabulated values of $\Omega_{x}$ and $H_{0}$ at $z=0$ are predictions assuming standard $\mathrm{\Lambda CDM}$, not direct measurements.

Since the values of PLANCK assume a vanishing pressure for dark matter, it is equivalent to saying that it is a model-independent measurement of dark matter density $\left.  \rho_{\mathrm{DM}}\right \vert _{\mathrm{CMB}}$ at $z=1089$. Using the value Eq.~(\ref{Eq_Omega_DM_PLANCK}), it is simple to check (omitting errors) that
\begin{align}
\left.  \rho_{\mathrm{DM}}\right \vert _{\mathrm{CMB}}^{\left(  \mathrm{PLANCK}\right)  } &  =\left.  \frac{3H_{0}^{2}\Omega_{\mathrm{DM}}^{\left( \mathrm{PLANCK}\right)  }}{\kappa_{\mathrm{4}}c^{2}}\left(  z+1\right)^{3}\right \vert_{z=1089}\nonumber \\
&  =0.262\, \frac{\mathrm{J}}{\mathrm{m}^{3}}\,.
\end{align}

In the context of our model, measurements of Riemannian geometry-dependent features (as CMB data) are sensitive only to the whole dressed effective dark matter density $\rho_{\mathrm{eff}}$, and not to the bare one $\rho_{\mathrm{DM}}$. Therefore, PLANCK data corresponds in our model to
\begin{equation}
\left.  \rho_{\mathrm{eff}}\right \vert _{z=1089}=\left.  \rho_{\mathrm{DM}}\right \vert _{\mathrm{CMB}}^{\left(  \mathrm{PLANCK}\right)  }\,,
\end{equation}
and not to the bare density $\left.  \rho_{\mathrm{DM}}\right \vert _{z=1089}$.

Even further, assuming that direct measurements give the correct value of $H_{0}$, like the ones from Refs.~\cite{Shajib:2019toy,Riess:2019cxk}, (e.g. $H_{0\mathrm{dir}}=h_{\mathrm{dir}}\, \frac{100\, \mathrm{km}}{\mathrm{s}\cdot \mathrm{Mpc}}$ with $h_{\mathrm{dir}}=0.742\pm0.018$), and allowing for a small negative effective barotropic constant $-1/3<\omega_{\mathrm{eff}}\leq 0$, it is clear that the values of $\Omega_{\mathrm{b}}$ and $\Omega_{\mathrm{eff}}$ slightly change in comparison to the tabulated $\Omega_{\mathrm{b}}^{\left(  \mathrm{PLANCK}\right)  }=0.049$ and $\Omega_{\mathrm{DM}}^{\left(  \mathrm{PLANCK}\right)  }=0.265$. For a start, the baryonic density parameter would correspond to
\begin{align}
\Omega_{\mathrm{b}} &  =\frac{\left(  \Omega_{\mathrm{b}}^{\left( \mathrm{PLANCK}\right)  }h^{2}\right)  _{\mathrm{tabulated}}}{h_{\mathrm{dir}}^{2}}\nonumber \\
&  =4.07\times10^{-2}.
\end{align}

Similarly, the parameter $\Omega_{\mathrm{eff}}$ depends on $\omega_{\mathrm{eff}}$ through
\begin{align}
\Omega_{\mathrm{eff}} &  =\left.  \frac{\kappa_{\mathrm{4}}c^{2}}{3H_{\mathrm{dir}0}^{2}\left(  z+1\right)  ^{3\left(  1+\omega_{\mathrm{eff}}\right)  }}\left.  \rho_{\mathrm{DM}}\right \vert _{\mathrm{CMB}}^{\left( \mathrm{PLANCK}\right)  }\right \vert _{z=1089},\nonumber \\
&  =\left.  \frac{\left(  \Omega_{\mathrm{DM}}^{\left(  \mathrm{PLANCK}\right)  }h^{2}\right)  _{\mathrm{tabulated}}}{h_{\mathrm{dir}}^{2}}\frac
{1}{\left(  z+1\right)  ^{3\omega_{\mathrm{eff}}}}\right \vert _{z=1089},\nonumber \\
&  =\frac{0.218}{1090^{3\omega_{\mathrm{eff}}}}.
\end{align}

From the flat spatial section geometry condition $\Omega_{\mathrm{eff}}+\Omega_{\mathrm{b}}+\Omega_{\Lambda}=1$, we have $\Omega_{\mathrm{eff}}=0.273$. Therefore, $\omega_{\mathrm{eff}}$ has the same small negative value already known from Ref.~\cite{Poplawski:2019tub},
\begin{equation}
\omega_{\mathrm{eff}}=-1.08\times10^{-2}\,.
\end{equation}

Using Eq.~(\ref{Eq_w_eff}), we can read back $\alpha_{\mathrm{Y}}$ as
\begin{align}
\alpha_{\mathrm{Y}}  & =\sqrt{-\frac{1}{1+\frac{1}{3\omega_{\mathrm{eff}}}}}\nonumber\\
& =0.183\,.
\end{align}

Therefore, from Eqs.~(\ref{Eq_w_eff}) and~(\ref{Eq_Torsiotropic}), we can see that torsion amplifies the dark matter density for only a small factor,
\begin{align}
\rho_{\mathrm{eff}}  & =\left(  1+\alpha_{\mathrm{Y}}^{2}\right) \rho_{\mathrm{DM}}\nonumber\\
& =1.03\,\rho_{\mathrm{DM}}\,,
\end{align}
and the \textbf{bare} torsion density parameter $\Omega_{\mathrm{DM}}$ corresponds finally to
\begin{align}
\Omega_{\mathrm{DM}}  & =\frac{1}{1.03}\Omega_{\mathrm{eff}}\nonumber \\
& =0.265\,,
\end{align}
which coincides precisely with the one tabulated by PLANCK. Therefore, the departure from $\mathrm{\Lambda CDM}$ is tiny. The only density parameter changing a bit is the one associated with baryons, decreasing from $\Omega_{\mathrm{b}}^{\left(  \mathrm{\Lambda CDM}\right)  }=0.049$ to $\Omega_{\mathrm{b}}=0.041$. The difference is due to the energy density associated with $\frac{3}{c^{2}\kappa_{\mathrm{4}}}Y^{2}$.

\section{Conclusions}

We have proved that a nonvanishing spin tensor for cold dark matter mimics the effect of a small negative barotropic constant, $\omega_{\mathrm{eff}}=-1.08\times10^{-2}$. The small negative pressure created through this mechanism suffices to explain the Hubble parameter tension, starting with the CMB initial conditions measured by PLANCK. The work of Ref.~\cite{Poplawski:2019tub} already proved that this small negative barotropic constant could solve the tension problem, but the novelty of the mechanism shown here is that it does not require \textquotedblleft exotic\textquotedblright \ physics. Instead, it only requires allowing one of the dark matter spin tensor components to have a non-zero value.

The departure from $\mathrm{\Lambda CDM}$ is tiny, and everything is in agreement with CMB's initial conditions. The only density parameter that changes a bit in comparison to standard $\mathrm{\Lambda CDM}$ is the one of baryonic matter, lowering from the standard $\mathrm{\Lambda CDM}$ value $0.049$ to $0.041$. The difference is due to the small energy density $\frac{3}{c^{2}\kappa_{\mathrm{4}}}Y^{2}$ associated with the spin. The tiny negative pressure corresponding to this spin component solves the tension problem in the current model.

To use a negligible spin tensor for baryonic matter is an excellent approximation  on cosmological scales, due to its interactions and decoherence. In contrast, none of these conditions seem to be true for dark matter, and therefore to neglect its spin tensor seems unjustified. That is the main reason that makes plausible the mechanism presented here.

Nevertheless, important unanswered questions remain when considering torsion as an extra dark source for the Riemannian piece of the geometry. Dark matter density is several orders higher in galaxies than the average in the Universe. Therefore, if torsion solves the Hubble parameter tension, it should also play a role in galaxy dynamics and formation of structures.

However, given our current knowledge (or lack of it), it seems extremely difficult to distinguish between the effects of \textquotedblleft bare\textquotedblright\ dark matter and the effects of torsion. Since torsion is dark, all of our observations are sensitive only to the combined effect of both phenomena. In cosmological scales, the requirements of homogeneity and isotropy are so strong that they fix the spin tensor almost wholly. It was vital to arrive at a model with just one free small parameter that could solve the Hubble tension. For cases with fewer symmetries, as galaxy dynamics and formation of structures, the same procedure is a non-starter. A priori, in our simple model, with a non-propagating torsion that follows the bare dark matter distribution, seems there are no reasons to expect a departure of $\Lambda\mathrm{CDM}$ concerning structure formation. Nonetheless, the only way to make a meaningful prediction would be to have a well-known dark matter Lagrangian instead of the simple model we have shown here.


The real way of solving this issue would be to have an experimental probe to distinguish between both effects. From the side of particle physics, the situation seems unclear in the foreseeable future. Since we have not detected dark matter particles, measuring their spin tensor is equivalent to measuring torsion in a particle physics experiment, an unsolved problem known for its enormous experimental difficulty~\cite{Puetzfeld:2014sja}. However, there is an entirely different way to attack this problem, which could be much more promising. Recently, Refs.~\cite{Barrientos:2019msu,Barrientos:2019awg} showed that a nonvanishing torsion background should affect the propagation of the amplitude and the polarization of gravitational waves. Therefore, gravitational waves could provide an astrophysical probe to test these ideas experimentally and check whether torsion could play a role in the dark matter puzzle.

\section{Acknowledgements}
FI is grateful to Antonella Cid, Pedro Labra\~{n}a, Perla Medina, Julio Oliva, and Jorge Zanelli for many enlightening discussions. FI is also thankful to Jos\'{e} Manuel Izquierdo for enlightening discussions and his warm hospitality at the University of Valladolid, were part of this work was done. OV acknowledges VRIIP-UNAP for financial support through Project VRIIP0062-19, and   FI acknowledges financial support from the Chilean government through FONDECYT grant 1180681 of the Government of Chile. FI is thankful of the emotional support of the Netherlands Bach Society. They made freely available superb quality recordings of the music of Bach. Without them, the 2020 pandemic quarantine would have been
unbearable, and this work, impossible.

\bibliographystyle{utphys}
\bibliography{Bib_7_jul_2020}

\providecommand{\href}[2]{#2}\begingroup\raggedright\begin{thebibliography}{10}

\bibitem{Lucca:2020zjb}
M.~Lucca and D.~C. Hooper, ``{Tensions in the dark: shedding light on Dark
  Matter-Dark Energy interactions},''
  \href{http://arxiv.org/abs/2002.06127}{{\ttfamily arXiv:2002.06127
  [astro-ph.CO]}}.

\bibitem{Alestas:2020mvb}
G.~Alestas, L.~Kazantzidis, and L.~Perivolaropoulos, ``{$H_0$ Tension, Phantom
  Dark Energy and Cosmological Parameter Degeneracies},''
  \href{http://arxiv.org/abs/2004.08363}{{\ttfamily arXiv:2004.08363
  [astro-ph.CO]}}.

\bibitem{Vattis:2019efj}
K.~Vattis, S.~M. Koushiappas, and A.~Loeb, ``{Dark matter decaying in the late
  Universe can relieve the H0 tension},''
  \href{http://dx.doi.org/10.1103/PhysRevD.99.121302}{{\em Phys. Rev. D}
  {\bfseries 99} no.~12, (2019) 121302},
  \href{http://arxiv.org/abs/1903.06220}{{\ttfamily arXiv:1903.06220
  [astro-ph.CO]}}.

\bibitem{Guo:2018ans}
R.-Y. Guo, J.-F. Zhang, and X.~Zhang, ``{Can the $H_0$ tension be resolved in
  extensions to $\Lambda$CDM cosmology?},''
  \href{http://dx.doi.org/10.1088/1475-7516/2019/02/054}{{\em JCAP} {\bfseries
  02} (2019) 054}, \href{http://arxiv.org/abs/1809.02340}{{\ttfamily
  arXiv:1809.02340 [astro-ph.CO]}}.

\bibitem{DiValentino:2017iww}
E.~Di~Valentino, A.~Melchiorri, and O.~Mena, ``{Can interacting dark energy
  solve the $H_0$ tension?},''
  \href{http://dx.doi.org/10.1103/PhysRevD.96.043503}{{\em Phys. Rev. D}
  {\bfseries 96} no.~4, (2017) 043503},
  \href{http://arxiv.org/abs/1704.08342}{{\ttfamily arXiv:1704.08342
  [astro-ph.CO]}}.

\bibitem{Rossi:2019lgt}
M.~Rossi, M.~Ballardini, M.~Braglia, F.~Finelli, D.~Paoletti, A.~A.
  Starobinsky, and C.~Umiltà, ``{Cosmological constraints on post-Newtonian
  parameters in effectively massless scalar-tensor theories of gravity},''
  \href{http://dx.doi.org/10.1103/PhysRevD.100.103524}{{\em Phys. Rev. D}
  {\bfseries 100} no.~10, (2019) 103524},
  \href{http://arxiv.org/abs/1906.10218}{{\ttfamily arXiv:1906.10218
  [astro-ph.CO]}}.

\bibitem{Braglia:2020iik}
M.~Braglia, M.~Ballardini, W.~T. Emond, F.~Finelli, A.~E. Gumrukcuoglu,
  K.~Koyama, and D.~Paoletti, ``{A larger value for $H_0$ by an evolving
  gravitational constant},'' \href{http://arxiv.org/abs/2004.11161}{{\ttfamily
  arXiv:2004.11161 [astro-ph.CO]}}.

\bibitem{Alcaniz:2019kah}
J.~Alcaniz, N.~Bernal, A.~Masiero, and F.~S. Queiroz, ``{Light Dark Matter: A
  Common Solution to the Lithium and ${H_0}$ Problems},''
  \href{http://arxiv.org/abs/1912.05563}{{\ttfamily arXiv:1912.05563
  [astro-ph.CO]}}.

\bibitem{Jedamzik:2020krr}
K.~Jedamzik and L.~Pogosian, ``{Relieving the Hubble tension with primordial
  magnetic fields},'' \href{http://arxiv.org/abs/2004.09487}{{\ttfamily
  arXiv:2004.09487 [astro-ph.CO]}}.

\bibitem{Freedman:2020dne}
W.~L. Freedman, B.~F. Madore, T.~Hoyt, I.~S. Jang, R.~Beaton, M.~G. Lee,
  A.~Monson, J.~Neeley, and J.~Rich, ``{Calibration of the Tip of the Red Giant
  Branch (TRGB)},'' \href{http://arxiv.org/abs/2002.01550}{{\ttfamily
  arXiv:2002.01550 [astro-ph.GA]}}.

\bibitem{Zumalacarregui:2020cjh}
M.~Zumalacarregui, ``{Gravity in the Era of Equality: Towards solutions to the
  Hubble problem without fine-tuned initial conditions},''
  \href{http://arxiv.org/abs/2003.06396}{{\ttfamily arXiv:2003.06396
  [astro-ph.CO]}}.

\bibitem{Yang:2018euj}
W.~Yang, S.~Pan, E.~Di~Valentino, R.~C. Nunes, S.~Vagnozzi, and D.~F. Mota,
  ``{Tale of stable interacting dark energy, observational signatures, and the
  $H_0$ tension},'' \href{http://dx.doi.org/10.1088/1475-7516/2018/09/019}{{\em
  JCAP} {\bfseries 09} (2018) 019},
  \href{http://arxiv.org/abs/1805.08252}{{\ttfamily arXiv:1805.08252
  [astro-ph.CO]}}.

\bibitem{Vagnozzi:2019ezj}
S.~Vagnozzi, ``{New physics in light of the $H_0$ tension: an alternative
  view},'' \href{http://arxiv.org/abs/1907.07569}{{\ttfamily arXiv:1907.07569
  [astro-ph.CO]}}.

\bibitem{DiValentino:2019jae}
E.~Di~Valentino, A.~Melchiorri, O.~Mena, and S.~Vagnozzi, ``{Nonminimal dark
  sector physics and cosmological tensions},''
  \href{http://dx.doi.org/10.1103/PhysRevD.101.063502}{{\em Phys. Rev. D}
  {\bfseries 101} no.~6, (2020) 063502},
  \href{http://arxiv.org/abs/1910.09853}{{\ttfamily arXiv:1910.09853
  [astro-ph.CO]}}.

\bibitem{Hehl1980}
F.~W. Hehl, ``{Four Lectures on Poincar{\'e} Gauge Field Theory},'' in {\em
  {Cosmology and Gravitation: Spin, Torsion, Rotation, and Supergravity}},
  P.~G. Bergmann and V.~De~Sabbata, eds., pp.~5--62.
\newblock Plenum Press, New York, 1980.
\newblock {Proceedings of the NATO Advanced Study Institute on Cosmology and
  Gravitation: Spin, Torsion, Rotation, and Supergravity, held at the Ettore
  Majorana International Center for Scientific Culture, Erice, Italy, May 6--8,
  1979.}

\bibitem{Blagojevic:2013xpa}
M.~Blagojevi{\'c} and F.~W. Hehl, eds., {\em {Gauge Theories of Gravitation}}.
\newblock World Scientific, Singapore, 2013.
\newblock
\url{https://doi.org/10.1142/p781}.
\newblock

\bibitem{Chakrabarty:2018ybk}
S.~Chakrabarty and A.~Lahiri, ``{Different types of torsion and their effect on
  the dynamics of fields},''
  \href{http://dx.doi.org/10.1140/epjp/i2018-12070-6}{{\em Eur. Phys. J. Plus}
  {\bfseries 133} no.~6, (2018) 242},
  \href{http://arxiv.org/abs/1907.02341}{{\ttfamily arXiv:1907.02341 [gr-qc]}}.

\bibitem{Alexander:2019wne}
S.~Alexander, M.~Cort\^es, A.~R. Liddle, J.~Magueijo, R.~Sims, and L.~Smolin,
  ``{The cosmology of minimal varying Lambda theories},''
\href{http://arxiv.org/abs/1905.10382}{{\ttfamily arXiv:1905.10382 [gr-qc]}}.

\bibitem{Magueijo:2019vmk}
J.~Magueijo and T.~Z{\l}o{\'s}nik, ``{Parity violating Friedmann Universes},''
  \href{http://dx.doi.org/10.1103/PhysRevD.100.084036}{{\em Phys. Rev. D}
  {\bfseries 100} no.~8, (2019) 084036},
  \href{http://arxiv.org/abs/1908.05184}{{\ttfamily arXiv:1908.05184 [gr-qc]}}.

\bibitem{Barker:2020gcp}
W.~Barker, A.~Lasenby, M.~Hobson, and W.~Handley, ``{Addressing $H_0$ tension
  with emergent dark radiation in unitary gravity},''
  \href{http://arxiv.org/abs/2003.02690}{{\ttfamily arXiv:2003.02690 [gr-qc]}}.

\bibitem{Soares-Santos:2019irc}
{\bfseries DES, LIGO Scientific, Virgo} Collaboration, M.~Soares-Santos {\em
  et~al.}, ``{First Measurement of the Hubble Constant from a Dark Standard
  Siren using the Dark Energy Survey Galaxies and the LIGO/Virgo
  Binary--Black-hole Merger GW170814},''
  \href{http://dx.doi.org/10.3847/2041-8213/ab14f1}{{\em Astrophys. J.}
  {\bfseries 876} no.~1, (2019) L7},
  \href{http://arxiv.org/abs/1901.01540}{{\ttfamily arXiv:1901.01540
  [astro-ph.CO]}}.

\bibitem{Poplawski:2019tub}
N.~Pop{\l}awski, ``{Non-particle dark matter from Hubble parameter},''
  \href{http://dx.doi.org/10.1140/epjc/s10052-019-7230-5}{{\em Eur. Phys. J. C}
  {\bfseries 79} no.~9, (2019) 734},
  \href{http://arxiv.org/abs/1906.03947}{{\ttfamily arXiv:1906.03947
  [physics.gen-ph]}}.

\bibitem{Poplawski:2013mra}
N.~J. Pop{\l}awski, ``{The energy and momentum of the Universe},''
  \href{http://dx.doi.org/10.1088/0264-9381/31/6/065005}{{\em Class. Quant.
  Grav.} {\bfseries 31} (2014) 065005},
  \href{http://arxiv.org/abs/1305.6977}{{\ttfamily arXiv:1305.6977 [gr-qc]}}.

\bibitem{Carroll:1994dq}
S.~M. Carroll and G.~B. Field, ``{Consequences of propagating torsion in
  connection dynamic theories of gravity},''
  \href{http://dx.doi.org/10.1103/PhysRevD.50.3867}{{\em Phys. Rev. D}
  {\bfseries 50} (1994) 3867--3873},
  \href{http://arxiv.org/abs/gr-qc/9403058}{{\ttfamily arXiv:gr-qc/9403058}}.

\bibitem{Chakrabarty:2019cau}
S.~Chakrabarty and A.~Lahiri, ``{Geometrical contribution to neutrino mass
  matrix},'' \href{http://dx.doi.org/10.1140/epjc/s10052-019-7209-2}{{\em Eur.
  Phys. J. C} {\bfseries 79} no.~8, (2019) 697},
  \href{http://arxiv.org/abs/1904.06036}{{\ttfamily arXiv:1904.06036
  [hep-ph]}}.

\bibitem{Kib61}
T.~W.~B. Kibble, ``Lorentz invariance and the gravitational field,''
  \href{http://dx.doi.org/10.1063/1.1703702}{{\em J. Math. Phys.} {\bfseries 2}
  (1961) 212--221}.

\bibitem{Sciama:1964wt}
D.~W. Sciama, ``{The Physical structure of general relativity},''
  \href{http://dx.doi.org/10.1103/RevModPhys.36.1103}{{\em Rev. Mod. Phys.}
  {\bfseries 36} (1964) 463--469}.
[Erratum: Rev. Mod. Phys. 36, 1103 (1964)].

\bibitem{Hehl:1971qi}
F.~W. Hehl and B.~K. Datta, ``{Nonlinear spinor equation and asymmetric
  connection in general relativity},''
\href{http://dx.doi.org/10.1063/1.1665738}{{\em J. Math. Phys.} {\bfseries 12}
  (1971) 1334--1339}.

\bibitem{doi:10.1142/6742}
H.~Kleinert, \href{http://dx.doi.org/10.1142/6742}{{\em Multivalued Fields}}.
\newblock World Scientific, 2008.
\newblock \url{https://doi.org/10.1142/6742}.

\bibitem{doi:10.1142/0356}
H.~Kleinert, \href{http://dx.doi.org/10.1142/0356}{{\em Gauge Fields in
  Condensed Matter}}.
\newblock World Scientific, 1989.
\newblock \url{https://doi.org/10.1142/0356}.

\bibitem{Hehl76}
F.~W. Hehl, P.~von~der Heyde, G.~D. Kerlick, and J.~M. Nester, ``General
  relativity with spin and torsion: Foundations and prospects,''
  \href{http://dx.doi.org/10.1103/RevModPhys.48.393}{{\em Rev. Mod. Phys.}
  {\bfseries 48} (1976) 393--416}.
  \url{http://link.aps.org/doi/10.1103/RevModPhys.48.393}.

\bibitem{Shapiro:2001rz}
I.~L. Shapiro, ``{Physical aspects of the space-time torsion},''
  \href{http://dx.doi.org/10.1016/S0370-1573(01)00030-8}{{\em Phys. Rept.}
  {\bfseries 357} (2002) 113},
\href{http://arxiv.org/abs/hep-th/0103093}{{\ttfamily arXiv:hep-th/0103093
  [hep-th]}}.

\bibitem{Hammond:2002rm}
R.~T. Hammond, ``{Torsion gravity},''
\href{http://dx.doi.org/10.1088/0034-4885/65/5/201}{{\em Rept. Prog. Phys.}
  {\bfseries 65} (2002) 599--649}.

\bibitem{Poplawski:2009fb}
N.~J. Pop{\l}awski, ``{Spacetime and Fields},''
\href{http://arxiv.org/abs/0911.0334}{{\ttfamily arXiv:0911.0334 [gr-qc]}}.

\bibitem{Valdivia:2017sat}
J.~Barrientos, F.~Cordonier-Tello, F.~Izaurieta, P.~Medina, D.~Narbona,
  E.~Rodríguez, and O.~Valdivia, ``{Nonminimal couplings, gravitational waves,
  and torsion in Horndeski's theory},''
  \href{http://dx.doi.org/10.1103/PhysRevD.96.084023}{{\em Phys. Rev.}
  {\bfseries D96} no.~8, (2017) 084023},
\href{http://arxiv.org/abs/1703.09686}{{\ttfamily arXiv:1703.09686 [gr-qc]}}.

\bibitem{Cid:2017wtf}
A.~Cid, F.~Izaurieta, G.~Leon, P.~Medina, and D.~Narbona, ``{Non-minimally
  coupled scalar field cosmology with torsion},''
  \href{http://dx.doi.org/10.1088/1475-7516/2018/04/041}{{\em JCAP} {\bfseries
  1804} no.~04, (2018) 041},
\href{http://arxiv.org/abs/1704.04563}{{\ttfamily arXiv:1704.04563 [gr-qc]}}.

\bibitem{Barrientos:2019awg}
J.~Barrientos, F.~Cordonier-Tello, C.~Corral, F.~Izaurieta, P.~Medina,
  E.~Rodríguez, and O.~Valdivia, ``{Luminal Propagation of Gravitational Waves
  in Scalar-tensor Theories: The Case for Torsion},''
  \href{http://dx.doi.org/10.1103/PhysRevD.100.124039}{{\em Phys. Rev. D}
  {\bfseries 100} no.~12, (2019) 124039},
  \href{http://arxiv.org/abs/1910.00148}{{\ttfamily arXiv:1910.00148 [gr-qc]}}.

\bibitem{Alexander:2020umk}
S.~Alexander, L.~Jenks, P.~Jirouvsek, J.~Magueijo, and T.~Z{\l}o{\'s}nik,
  ``{Gravity waves in parity-violating Copernican Universes},''
  \href{http://arxiv.org/abs/2001.06373}{{\ttfamily arXiv:2001.06373 [gr-qc]}}.

\bibitem{Barrientos:2019msu}
J.~Barrientos, F.~Izaurieta, E.~Rodríguez, and O.~Valdivia, ``{Wave Operators,
  Torsion, and the Weitzenb\"{o}ck Identities},''
\href{http://arxiv.org/abs/1903.04712}{{\ttfamily arXiv:1903.04712 [gr-qc]}}.

\bibitem{Puetzfeld:2014sja}
D.~Puetzfeld and Y.~N. Obukhov, ``{Prospects of detecting spacetime torsion},''
  \href{http://dx.doi.org/10.1142/S0218271814420048}{{\em Int. J. Mod. Phys. D}
  {\bfseries 23} no.~12, (2014) 1442004},
  \href{http://arxiv.org/abs/1405.4137}{{\ttfamily arXiv:1405.4137 [gr-qc]}}.

\bibitem{SupergravityVanProeyen}
D.~Z. Freedman and A.~V. Proeyen, {\em Supergravity}.
\newblock Cambridge University Press, 1~ed., 5, 2012.

\bibitem{Weyssenhoff:1947iua}
J.~Weyssenhoff and A.~Raabe, ``{Relativistic dynamics of spin-fluids and
  spin-particles},'' {\em Acta Phys. Polon.} {\bfseries 9} (1947) 7--18.

\bibitem{Obukhov:1987yu}
Y.~Obukhov and V.~Korotkii, ``{The Weyssenhoff fluid in Einstein-Cartan
  theory},'' \href{http://dx.doi.org/10.1088/0264-9381/4/6/021}{{\em Class.
  Quant. Grav.} {\bfseries 4} (1987) 1633--1657}.

\bibitem{Boehmer:2006gd}
C.~G. Boehmer and P.~Bronowski, ``{The Homogeneous and isotropic Weyssenhoff
  fluid},'' {\em Ukr. J. Phys.} {\bfseries 55} no.~5, (2010) 607--612,
  \href{http://arxiv.org/abs/gr-qc/0601089}{{\ttfamily arXiv:gr-qc/0601089}}.

\bibitem{Brechet:2008zz}
S.~Brechet, M.~Hobson, and A.~Lasenby, ``{Classical big-bounce cosmology:
  Dynamical analysis of a homogeneous and irrotational Weyssenhoff fluid},''
  \href{http://dx.doi.org/10.1088/0264-9381/25/24/245016}{{\em Class. Quant.
  Grav.} {\bfseries 25} (2008) 245016},
  \href{http://arxiv.org/abs/0807.2523}{{\ttfamily arXiv:0807.2523 [gr-qc]}}.

\bibitem{Poplawski:2011jz}
N.~J. Pop{\l}awski, ``{Nonsingular, big-bounce cosmology from spinor-torsion
  coupling},'' \href{http://dx.doi.org/10.1103/PhysRevD.85.107502}{{\em Phys.
  Rev. D} {\bfseries 85} (2012) 107502},
  \href{http://arxiv.org/abs/1111.4595}{{\ttfamily arXiv:1111.4595 [gr-qc]}}.

\bibitem{Poplawski:2010kb}
N.~J. Pop{\l}awski, ``{Cosmology with torsion: An alternative to cosmic
  inflation},'' \href{http://dx.doi.org/10.1016/j.physletb.2010.09.056}{{\em
  Phys. Lett. B} {\bfseries 694} (2010) 181--185},
  \href{http://arxiv.org/abs/1007.0587}{{\ttfamily arXiv:1007.0587
  [astro-ph.CO]}}. [Erratum: Phys.Lett.B 701, 672--672 (2011)].

\bibitem{Poplawski:2014dea}
N.~Pop{\l}awski, ``{Universe in a Black Hole in Einstein--cartan Gravity},''
  \href{http://dx.doi.org/10.3847/0004-637X/832/2/96}{{\em Astrophys. J.}
  {\bfseries 832} no.~2, (2016) 96},
  \href{http://arxiv.org/abs/1410.3881}{{\ttfamily arXiv:1410.3881 [gr-qc]}}.

\bibitem{Poplawski:2018ypb}
N.~Pop{\l}awski, ``{The simplest origin of the big bounce and inflation},''
  \href{http://dx.doi.org/10.1142/S021827181847020X}{{\em Int. J. Mod. Phys. D}
  {\bfseries 27} no.~14, (2018) 1847020},
  \href{http://arxiv.org/abs/1801.08076}{{\ttfamily arXiv:1801.08076
  [physics.pop-ph]}}.

\bibitem{Boehmer:2008ah}
C.~G. Boehmer and J.~Burnett, ``{Dark spinors with torsion in cosmology},''
  \href{http://dx.doi.org/10.1103/PhysRevD.78.104001}{{\em Phys. Rev. D}
  {\bfseries 78} (2008) 104001},
  \href{http://arxiv.org/abs/0809.0469}{{\ttfamily arXiv:0809.0469 [gr-qc]}}.

\bibitem{Tilquin:2011bu}
A.~Tilquin and T.~Schucker, ``{Torsion, an alternative to dark matter?},''
  \href{http://dx.doi.org/10.1007/s10714-011-1222-6}{{\em Gen. Rel. Grav.}
  {\bfseries 43} (2011) 2965--2978},
  \href{http://arxiv.org/abs/1104.0160}{{\ttfamily arXiv:1104.0160
  [astro-ph.CO]}}.

\bibitem{Aghanim:2018eyx}
{\bfseries Planck} Collaboration, N.~Aghanim {\em et~al.}, ``{Planck 2018
  results. VI. Cosmological parameters},''
  \href{http://arxiv.org/abs/1807.06209}{{\ttfamily arXiv:1807.06209
  [astro-ph.CO]}}.

\bibitem{Shajib:2019toy}
{\bfseries DES} Collaboration, A.~Shajib {\em et~al.}, ``{STRIDES: a 3.9 per
  cent measurement of the Hubble constant from the strong lens system DES
  J0408-5354},'' \href{http://arxiv.org/abs/1910.06306}{{\ttfamily
  arXiv:1910.06306 [astro-ph.CO]}}.

\bibitem{Riess:2019cxk}
A.~G. Riess, S.~Casertano, W.~Yuan, L.~M. Macri, and D.~Scolnic, ``{Large
  Magellanic Cloud Cepheid Standards Provide a 1\% Foundation for the
  Determination of the Hubble Constant and Stronger Evidence for Physics beyond
  $\Lambda$CDM},'' \href{http://dx.doi.org/10.3847/1538-4357/ab1422}{{\em
  Astrophys. J.} {\bfseries 876} no.~1, (2019) 85},
  \href{http://arxiv.org/abs/1903.07603}{{\ttfamily arXiv:1903.07603
  [astro-ph.CO]}}.

\end{thebibliography}\endgroup

\end{document}